\documentclass[runningheads]{llncs}
\usepackage{paralist}
\usepackage{subcaption}
\usepackage{hyperref}
\hypersetup{
colorlinks = true,
linkcolor=black,
citecolor=black,
urlcolor=black
}

\usepackage{microtype}
\usepackage{booktabs}
\usepackage{verbatim}
\usepackage{makecell}
\usepackage{xspace}
\usepackage[T1]{fontenc}
\usepackage{graphicx}
\usepackage{tabularx}
\usepackage{amsmath}
\usepackage{multicol}
\usepackage{multirow}
\usepackage[most]{tcolorbox}
\usepackage{cleveref}
\usepackage{float}
\usepackage{placeins}
\crefname{equation}{eq.}{eqs.}
\Crefname{equation}{Eq.}{Eqs.}

\newcommand{\none}{\texttt{-}\xspace}
\newcommand{\neglibleStrength}{\texttt{negligible}\xspace}
\newcommand{\smallStrength}{\texttt{small}\xspace}
\newcommand{\mediumStrength}{\texttt{medium}\xspace}
\newcommand{\largeStrength}{\texttt{large}\xspace}
\newcommand{\noDifference}{\ensuremath{\equiv}\xspace}

\newcommand{\clip}{\texttt{CLIP}\xspace}
\newcommand{\blip}{\texttt{BLIP}\xspace}

\newcommand{\ourApproach}{\texttt{MetaVLM}\xspace}
\newcommand{\randomSearch}{\texttt{Rand}\xspace}
\newcommand{\Atwelve}{\ensuremath{\hat{A}}\textsubscript{12}\xspace}
\newcommand{\vlm}{\ensuremath{\mathcal{V}}\xspace}
\newcommand{\prompt}{\ensuremath{P}\xspace}
\newcommand{\image}{\ensuremath{I}\xspace}
\newcommand{\transImage}{\ensuremath{\image^{\prime}}\xspace}
\newcommand{\genericImage}{\ensuremath{X}\xspace}
\newcommand{\trans}{\ensuremath{T}\xspace}
\newcommand{\transParam}{\ensuremath{\theta}\xspace}
\newcommand{\vlmRes}{\ensuremath{y}\xspace}
\newcommand{\vlmTranRes}{\ensuremath{y^\prime}\xspace}
\newcommand{\setMRs}{\ensuremath{\mathcal{M}}\xspace}
\newcommand{\fPred}{\ensuremath{\mathit{f_{pred}}}\xspace}
\newcommand{\fTrans}{\ensuremath{\mathit{f_{trans}}}\xspace}
\newcommand{\countPre}{\ensuremath{\mathit{Ccount}}\xspace}
\newcommand{\countPreOneArg}[1]{\ensuremath{\countPre_{#1}}\xspace}
\newcommand{\countPreTwoArgs}[2]{\ensuremath{\countPre_{#1}^{#2}}\xspace}
\newcommand{\Cconf}{\ensuremath{\mathit{Cconf}}\xspace}
\newcommand{\CconfOneArg}[1]{\ensuremath{\Cconf_{#1}}\xspace}
\newcommand{\CconfTwoArgs}[2]{\ensuremath{\Cconf_{#1}^{#2}}\xspace}
\newcommand{\partner}{\texttt{DNV AS}\xspace}
\newcommand{\EUProject}{\texttt{InnoGuard}\xspace}
\newcommand{\thBVclip}{\texttt{Th}\xspace}
\newcommand{\BV}{\ensuremath{\mathrm{BV}}\xspace}
\newcommand{\BVblip}{\ensuremath{\BV_{\mathrm{blip}}}\xspace}
\newcommand{\BVclip}{\ensuremath{\BV_{\mathrm{clip}}}\xspace}
\newcommand{\violationStrength}{\ensuremath{\mathrm{VS}\xspace}}
\newcommand{\VSblip}{\ensuremath{\violationStrength_{\mathrm{blip}}}\xspace}
\newcommand{\VSclip}{\ensuremath{\violationStrength_{\mathrm{clip}}}\xspace}
\newcommand{\benchSet}{\ensuremath{\mathit{Bench}}\xspace}
\newcommand{\benchImage}[1]{\ensuremath{\mathit{B}_{#1}}\xspace}
\newcommand{\overTh}{\ensuremath{\mathit{violClasses}}\xspace}
\newcommand{\transParamMax}{\ensuremath{\transParam^{\mathit{max}}}\xspace}
\newcommand{\violRate}{\ensuremath{\mathit{VR}}\xspace}
\newcommand{\selectedMRs}{\ensuremath{\mathit{SelMRsInds}}\xspace}
\newcommand{\allViolations}{\ensuremath{A}\xspace}
\newcommand{\mr}{\text{MR}\xspace}
\newcommand{\mrNum}[1]{\text{MR}#1\xspace}
\newcommand{\mrs}{\text{MRs}\xspace}
\newcommand{\mrComposition}{MR com\-po\-si\-tion\xspace}
\newcommand{\mrCompositions}{MR com\-po\-si\-tions\xspace}
\newcommand{\composition}{\text{Com.}\xspace}
\newcommand{\ord}{\text{Order}}
\newcommand{\solutionSet}{\ensuremath{\mathit{S}}\xspace}
\newcommand{\smallSolution}{\ensuremath{\mathit{g}}\xspace}
\newcommand{\violRatePerc}[1]{\ensuremath{\violRate_{\mathrm{#1}}}\xspace}
\newcommand{\violRateOne}{\ensuremath{\violRatePerc{1\%}}\xspace}
\newcommand{\violRateThree}{\ensuremath{\violRatePerc{3\%}}\xspace}
\newcommand{\violRateFive}{\ensuremath{\violRatePerc{5\%}}\xspace}
\newcommand{\violStrengthOne}{\ensuremath{\violationStrength_\mathrm{1\%}}\xspace}
\newcommand{\violStrengthThree}{\ensuremath{\violationStrength_{\mathrm{3\%}}}\xspace}
\newcommand{\violStrengthFive}{\ensuremath{\violationStrength_{\mathrm{5\%}}}\xspace}
\newcommand{\finding}[1]{\begin{tcolorbox}[colframe=black, width=1\linewidth, colback=gray!15, left=2pt,right=2pt,top=2pt,bottom=2pt, breakable]#1\end{tcolorbox}}

\makeatletter
\def\orcidID#1{\kern .08em\href{https://orcid.org/#1}{\includegraphics[keepaspectratio,width=0.9em]{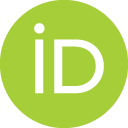}}}
\makeatother

\begin{document}

\title{Search-Based Metamorphic Testing of Vision-Language Models in Autonomous Underwater Robotic Software}

\author{Muhammad Yousaf\inst{1,2}\orcidID{0009-0003-7866-1933} \and
Aitor Arrieta\inst{3}\orcidID{0000-0001-7507-5080} \and
Shaukat Ali\inst{1}\orcidID{0000-0002-9979-3519} \and
Paolo Arcaini\inst{4}\orcidID{0000-0002-6253-4062} \and
Shuai Wang\inst{5}\orcidID{0000-0003-3164-7002}}
\authorrunning{M. Yousaf et al.}

\institute{Simula Research Laboratory, Oslo, Norway \and
Oslo Metropolitan University, Oslo, Norway \and
Mondragon University, Mondragon, Spain \and
National Institute of Informatics, Tokyo, Japan \and
Group Research and Development, DNV AS, Oslo, Norway\\ 
\email{muhammady@simula.no, aarrieta@mondragon.edu, shaukat@simula.no, arcaini@nii.ac.jp, shuai.wang@dnv.com}
}

\titlerunning{Search-Based Metamorphic Testing of VLMs in AUR Software}

\maketitle 

\begin{abstract}
Our industry partner focuses on quality assurance for industrial systems across multiple domains, including maritime systems, such as overwater vessels and autonomous underwater robots (AURs). Despite the strong performance of vision-language models (VLMs) in scene understanding, image captioning, and object recognition, their use in AUR software operating in underwater environments is underexplored. Therefore, in this context, it is important to evaluate the quality of VLMs for integration into AUR software and, so, automated software testing tools are needed to assess their suitability and improve their dependability. To this end, we propose a search-based metamorphic testing approach (\ourApproach) that identifies a minimal set of transformations on underwater images to induce incorrect model predictions, thereby revealing VLM failures. We employ NSGA-II as a multi-objective search algorithm and evaluate it over open-source VLMs \blip and \clip, against a random search baseline. Results demonstrate the strengths and limitations of each VLM in the context of AUR software systems. Based on the results, we derive lessons for software engineering practitioners and researchers working on quality assurance of VLM-based software systems.
\end{abstract}

\section{Introduction}\label{sec:introduction}
Autonomous underwater robots (AURs) are used in the maritime domain to perform a wide range of activities, such as underwater surveying, trash collection, and monitoring underwater equipment 
(e.g., in the oil and gas sector).
Such AURs often employ perception software that processes images of the underwater environment to analyze scenes and perform various tasks~\cite{UOD_survey_2025}. To this end, perception software often relies on deep learning models that, however, depend on scarce labeled data. Moreover, the underwater environment itself is challenging due to motion, turbidity, and illumination variations. Vision-language models (VLMs) have demonstrated promising performance across tasks such as object identification and scene understanding with limited or no labeled data~\cite{Alawode2025AquaticCLIP,vla_dualBrain_2025}. As a result, there is an increasing industrial interest in evaluating their performance for use as components within AUR software. Naturally, this calls for techniques to test VLMs, assess their suitability for critical applications, and further improve their quality to enable their use in AUR software.

Towards building systematic techniques for testing VLMs in AUR software, as part of the EU project \EUProject (\url{https://innoguard.eu/}), we are developing novel methods to ensure the dependability of autonomous cyber-physical systems, including AURs, with a focus on the use of AI foundation models, specifically VLMs in this paper. This work is carried out in collaboration with \partner{},  
which specializes in quality assurance and certification for industrial systems in different domains, including the maritime one. To this end, the focus of this paper is to develop a systematic, automated testing tool to assess the industrial use of VLMs in AUR software as a starting point, and later for other applications in the maritime domain.

We present \ourApproach, a multi-objective, search-based metamorphic testing approach for VLMs to reveal prediction failures in response to changes in underwater images. Metamorphic testing addresses the test oracle problem using metamorphic relations to identify incorrect VLM behavior. \ourApproach has two objectives:
\begin{inparaenum}[(i)]
\item applying a minimal sequence of transformations to a source image to ensure that it remains realistic;
\item maximizing the number of prediction failures induced by the differences between the source and the transformed image.
\end{inparaenum}

We employ NSGA-II as the multi-objective search algorithm in \ourApproach and test two open-source VLMs, \blip and \clip, on a set of underwater images to assess their robustness. We use random search (\randomSearch) as a comparison baseline. Our results show that \ourApproach is significantly more effective than \randomSearch in detecting VLM prediction failures. In particular, \ourApproach discovers low‑order \mrCompositions with higher violation rates than \randomSearch. Moreover, \ourApproach detected more severe violations than \randomSearch for both VLMs even with minimal transformations. 
These findings highlight the importance of multi-objective search in identifying not merely more failures, but failures with minimal image transformations. Finally, we provide a set of lessons learned from our empirical evaluation that are beneficial for other practitioners and researchers working on robotic software and interested in investigating the use of VLMs in their context.

\section{Industrial Context} \label{sec:industrcontex}
In the maritime domain, \partner operates as an independent third-party provider of classification, certification, and advisory services, with the overarching goal of safeguarding life, property, and the environment. As a classification society, the key objective is to establish rules and standards to verify that vessels and their associated systems comply with these requirements throughout their lifecycle, from design and construction to operation and modification.

With the rapid maritime digitalization paving the way, vessels are increasingly realized as complex cyber-physical systems composed of intelligent software subsystems (e.g., control and navigation). These subsystems are often developed by multiple vendors and integrated late in the lifecycle, introducing emerging system behaviors (e.g., interoperability) that are challenging to verify from an industrial assurance perspective. 
The assurance of complex maritime software systems increasingly depends on advanced techniques, such as simulation-based testing, to evaluate system behavior across diverse, safety-critical scenarios. At the same time, assurance is shifting toward continuous, data-driven processes, where large volumes of test and operational data are collected and used.

In this paper, we focus on Autonomous Underwater Robots (AURs) used for industrial tasks, such as inspection, underwater repair, and trash collection. We focus on underwater trash collection as a representative use case and test object detection capabilities enabled by VLMs in AURs. In maritime settings, reliable detection and classification of trash are essential for dependable operations, e.g., to avoid damaging vegetation. However, challenging conditions, such as low visibility and occlusion, degrade performance, making robust perception models critical for deployment and necessitating systematic, automated testing.
We collaborate with \partner to design a systematic, automated testing of VLMs in AURs to expose robustness issues and support the dependable integration of VLMs into AUR software.

\section{Proposed approach \ourApproach}\label{sec:approach}

\subsection{Approach Overview}\label{subsec: approach_overview}

We represent an AUR's VLM-based perception module as \vlm, which takes as input a textual prompt \prompt and an image \image, producing an output $\vlmRes=\vlm(\prompt, \image)$, where \vlmRes denotes the predicted labels with object counts associated with each predicted class. We reuse the prompt \prompt from~\cite{yousafICST2026arxiv}, which is an instruction-style, zero-shot, domain-specific prompt. \Cref{tab:prompt_template} reports the prompt template and an example output.
\begin{table}[!t]
\centering
\caption{Prompt template and example output}
\label{tab:prompt_template}
\resizebox{\columnwidth}{!}{
\scriptsize
\begin{tabular}{p{40pt}|p{200pt}|p{90pt}}
\toprule
& \textbf{Prompt Template} & \textbf{Example Output} \\
\midrule
System Message
& ``Identify and list all visible objects in this underwater image in four categories: (1) Animals, (2) Vegetation, (3) Objects, and (4) Trash. For each category, list the detected items with counts. If a category is definitely absent, write ‘None'.'' & (same for both VLMs) \\
\midrule
\multirow{4}{*}{Classes} & Animals: [detected animals with counts] & Animals: 1 fish \\
& Vegetation: [detected plants with counts] & Vegetation: 1 plant \\
& Objects: [man-made objects with counts] & Objects: 1 plastic bottle \\
& Trash: [trash items with counts] & Trash: 1 piece of trash \\
\midrule
\makecell[t]{Final\\Labels}
& Predicted classes & \{`trash', `object', `vegetation', `animal'\} \\
\bottomrule
\end{tabular}
}
\end{table}
We adopt six equality-based metamorphic relations (\mrs) based on geometric image properties used in~\cite{adaptive_mt_testing_OD_models,mt_optimize2024}, namely rotation, scaling, histogram equalization, downsampling, shear, and translation (see \Cref{tab:mr_mmr_details}).
\begin{table}[!t]
\centering
\caption{Equality-based \mrs applied in \ourApproach (adapted from~\cite{adaptive_mt_testing_OD_models})}
\label{tab:mr_mmr_details}
\scriptsize
\setlength{\tabcolsep}{6pt}
\begin{tabular}{lllcc}
\toprule
\textbf{\mr} & \textbf{Name} & \textbf{Parameter} & \textbf{Nominal value $\transParam^{n}$} & \textbf{Range of \transParam} \\
\midrule
\mrNum{1} & Rotation & Angle & $0^\circ$ & $[-5^\circ, +5^\circ]$ \\ \hline
\mrNum{2} & Scaling & Zoom factor & 1 & $[0.7, 1.3]$ \\ \hline
\mrNum{3} & Histogram eq. & Clip limit & 1 & $[1.0, 2.0]$ \\ \hline
\mrNum{4} & Downsampling & Scale factor & 1 & $[0.9, 1.0]$ \\ \hline
\mrNum{5} & Shear & Horizontal shear coefficient & 0 & $[-0.15, +0.15]$ \\ \hline
\mrNum{6} & Translation & Horizontal shift percentage & 0 & $[-0.03, +0.03]$ \\ 
\bottomrule
\end{tabular}
\end{table}
%
These \mrs, in the context of VLMs, have not been tested on underwater images before; however, they have been studied in other domains, e.g., autonomous driving~\cite{deepRoad2018}. A metamorphic transformation $\trans_{\transParam}$ is defined as a parameterized function applied to a source image \image, producing a transformed image \transImage as follows\footnote{For simplicity, in the paper we will talk about {\it source image} and {\it transformed image} instead of {\it source test case} and {\it follow-up test case}.}:
\[
\transImage = \trans_{\transParam}(\image)
\]
where \transParam denotes the transformation parameters. These \mrs capture realistic variations encountered in underwater environments, preserving as much semantic information as possible. The expected behavior of a robust \vlm is:
\[
\vlmRes = \vlmTranRes \qquad \text{with} \; \vlmRes = \vlm(\prompt, \image) \; \text{ and } \; \vlmTranRes = \vlm(\prompt, \transImage)
\]

Our goal is to apply minimal transformations that induce the maximum changes in predictions, thereby revealing \mr violations (i.e., $\vlmRes \neq \vlmTranRes$). We apply multiple \mrs (called {\it \mrCompositions}) over an image rather than a single one because real-world underwater images are often affected by several simultaneous factors, e.g., AUR motion may cause rotation, translation, and blur, while changing depth may also reduce brightness and contrast. In addition, \mrCompositions are well known to reduce testing costs~\cite{qiu2020theoretical,arrieta2022cost}. We select multiple \mrs and apply them sequentially to an image \image to obtain a transformed image \transImage. We call {\it order} (between 1 and 6) the number of applied \mrs within an \mrComposition.

We formulate the problem as a multi-objective optimization problem. For the selected \mrs and their transformation parameters \transParam (e.g., rotation angle for \mrNum{2} in \Cref{tab:mr_mmr_details}), we aim to
\begin{compactenum}[(1)]
\item minimize the magnitude of the transformations ($\trans_{\transParam}$) applied to the input as compared to nominal values of that parameter, and
\item maximize the VLMs predictions difference between \vlmRes and \vlmTranRes.
\end{compactenum}
The first objective ensures that we apply subtle, realistic changes, while the second objective captures the severity of the model’s prediction differences between the source and transformed image. The VLM’s prediction differences are quantified either using object counts (i.e., the difference in the number of objects identified in \image and \transImage) or prediction confidence scores (i.e., the difference in confidence scores associated with each class). We consider both measures because different VLMs may produce outputs in different formats.

\subsection{Solution Encoding} \label{subsec:sol_encoding}

The solution encoding defines which of the six \mrs are selected in \mrComposition and their associated parameters \transParam. \Cref{tab:list_mmrs} shows the solution encoding in the first row and an example of individual in the second row.
\begin{table}[!t]
\centering
\caption{General encoding of metamorphic relations}
\label{tab:list_mmrs}
\scriptsize
\setlength{\tabcolsep}{4pt}
\begin{tabular}{c|c|c|c|c|c|c|c|c|c|c|c|c|c|c}
\toprule
& \multicolumn{2}{c|}{\textbf{\mrNum{1} (Rotation)}} & \multicolumn{2}{|c|}{\textbf{\mrNum{2} (Scaling)}} & \multicolumn{8}{c|}{\dots} & \multicolumn{2}{|c}{\textbf{\mrNum{6} (Translation)}} \\
\midrule
Encoding & Bool & \transParam (angle) & Bool & \transParam (zoom) & \multicolumn{8}{c|}{\dots} & Bool & \transParam (shift) \\
\hline
Example & True & $3^\circ$ & False & 0.0 & \multicolumn{8}{c|}{\dots} & True & 0.02 \\
\bottomrule
\end{tabular}
\end{table}
The search space consists of 12 variables, organized into six pairs of activation and transformation parameter variables, each corresponding to one \mr. For each \mr, the first variable is a Boolean that indicates the status of the \mr, i.e., whether it is active or not; the second variable, instead, indicates its associated parameter value. \mr parameters for \mrNum{1}-\mrNum{6} are shown in \Cref{tab:mr_mmr_details}, where the \textit{Parameter} column specifies the parameter name, the \textit{Nominal value} column shows the default values, while the \textit{Range} column specifies the allowed range of parameter values. For example, for \mrNum{1} (Rotation), the parameter is the angle, with values ranging from -5.0 to +5.0.
When two or more \mrs are selected, they are applied sequentially in the fixed order. For example, if \mrNum{3} and \mrNum{6} are selected, \mrNum{3} will be applied to the input image \image to obtain \transImage, followed by the application of \mrNum{6} to \transImage to obtain the final transformed image (for simplicity, we call \transImage even the last obtained transformed image).

\subsection{Objective Functions} \label{subsec:objective_functions}

Given a source image \image and a transformed image \transImage, the evaluation with \vlm produces outputs \vlmRes and \vlmTranRes. The problem has two objectives:
\[
\min \fTrans \qquad \text{and} \qquad \max \fPred
\]

\subsubsection{Metamorphic Relation Magnitude}\label{subsec:MR_magnitude}
The first objective \fTrans minimizes the magnitude of the applied transformations, ensuring that the transformed image remains as close as possible to the source image. This is calculated as the normalized distance between the applied transformation parameters and their nominal values (see \Cref{tab:mr_mmr_details}). A smaller value means a more subtle transformation.

Let \setMRs denote the set of six \mrs, and \transParam denote the set of parameter values. \mrNum{$i$} denotes the $i$-th \mr, whereas $\transParam_i$ denotes the parameter value of the $i$-th \mr. A solution consists of a set of selected \mr indexes, denoted by \selectedMRs, and $\transParam_{i}^{n}$ denotes the normal parameter value of the $i$-th \mr, as specified in \Cref{tab:mr_mmr_details}. Based on the selected \mr indexes, we calculate the objective functions as:
\[
\fTrans =
\sum\limits_{j\in \selectedMRs} \frac{|\transParam_{j}^{n} - \transParam_j|}{\transParamMax_{j}} 
\]
%
where $\transParamMax_{j}$ represents the maximum deviation for the parameter of $j$-th \mr. For example, for zoom the maximum scaling deviation relative to the nominal value is 0.3 (i.e., $\max(1-0.7, 1.3-1)$).

\subsubsection{Prediction Deviation} \label{subsec: prediction_deviation}
The second objective maximizes the difference between the model-predicted classes on the source and transformed images. Let $C$ be the set of classes that a \vlm predicts. For example, in the case of an AUR responsible for trash collection, possible objects to identify using the VLM are: 
\begin{equation}
C = \{\text{Trash}, \text{Objects}, \text{Animal}, \text{Vegetation}\}
\label{eq:target_classes}
\end{equation}

Due to differences in their architectures, some VLMs predict object counts, while other VLMs predict classes together with confidence scores. Considering those adopted in our experiments, \blip belongs to the former type, while \clip to the latter type. Therefore, depending on the nature of outputs, we define two different ways to compute this objective. For \blip, the objective is computed using the predicted object counts for each class it outputs. Let the number of occurrences of each predicted class for an image \genericImage by a \vlm be denoted as \countPreOneArg{\genericImage}. The count of the $i$-th predicted class is represented as \countPreTwoArgs{\genericImage}{i}. Given an image \image and the corresponding transformed image \transImage, the objective is computed as:
%
\begin{equation}
\label{eq:blip_objective}
\fPred = \sum_{j=1}^{|C|} \mathrm{nor}(|\countPreTwoArgs{\image}{j} - \countPreTwoArgs{\transImage}{j}|)
\quad \text{with} \;\mathrm{nor}(x) = \frac{x}{x+1}
\end{equation}

If a VLM returns confidence scores for each class (e.g., \clip), we define the prediction difference as the absolute difference in confidence scores between the source and transformed images. Let the confidence scores associated with the predicted classes on an image \genericImage by a \vlm be denoted by \CconfOneArg{\genericImage}, whereas the confidence of the $i$-th predicted class is represented as \CconfTwoArgs{\genericImage}{i}. Given an image \image and the corresponding transformed image \transImage, the objective is computed as:
\begin{equation}
\label{eq:clip_objective}
\fPred = \sum_{j=1}^{|C|}
\left|
\CconfTwoArgs{\image}{j}
-
\CconfTwoArgs{\transImage}{j}
\right|
\end{equation}

\section{Experimental Setup}\label{sec:setup}

We aim to assess whether \ourApproach can effectively expose failures in VLMs integrated into AUR software. To this end, we focus on the trash collection functionality of an AUR, in which the robot relies on its perception software to identify and collect waste. In this context, we test a VLM that performs such identification. Code and experimental results are available online: (\url{https://github.com/Myusuf121/MetaVLM}).

\subsection{Research Questions} \label{subsec:RQs}
We aim to evaluate the effectiveness of \ourApproach relative to a simple baseline, i.e., Random Search (\randomSearch), as well as study the effect of various \mrs and their characteristics on failures. Thus, we formulate the following research questions.
\begin{compactdesc}
\item \textbf{RQ1}: {\it How is the effectiveness of \ourApproach compared to that of \randomSearch?}\\
We investigate whether \ourApproach, with NSGA-II as search algorithm, is necessary to identify \mr violations, or a simple approach like \randomSearch is sufficient. 

\item \textbf{RQ2}: {\it Which \mrCompositions are effective in triggering VLM prediction failures?}
We answer this RQ with two sub-RQs.
\begin{compactdesc}

\item \textbf{RQ2.a} {\it How do the order of \mrCompositions relate to prediction violations?}

We study how the order of \mrCompositions (1 to 6) relates to prediction violations, assessed using a binary metric. Such an assessment depends on the output produced by a VLM. For example, a violation occurs if at least one main label changes in the transformed image compared to the source image.
\item \textbf{RQ2.b} {\it Which exact \mrCompositions achieve the highest number of violations?}

Compared to RQ2.a, we analyze specific \mrCompositions rather than their occurrences, e.g., \{\mrNum{2}, \mrNum{3}, \mrNum{4}{}\}, indicating these \mrs were applied to the input image in this order. 
\end{compactdesc}
\item \textbf{RQ3}: {\it What is the strength of violations induced by the \mrCompositions?}

We study the strength of violation by measuring the extent to which the outputs differ between the source and transformed images.

\end{compactdesc}

\subsection{Subject Images and VLMs}\label{subsec:subject_images}
For our experiments, we selected images from the SeaClear dataset~\cite{seaclear_nature}, an open-source underwater dataset containing high-resolution (1920 $\times$ 1080) images captured in diverse environments, including pools, oceans, and deep-sea. From this dataset, we selected images correctly classified by \blip~\cite{Li2022BLIP} (a VLM that showed superior performance in~\cite{yousafICST2026arxiv}), as our goal is to detect \mr violations rather than existing classification errors. 
Thus, the final set \benchSet contains 30 correctly classified underwater images covering 5 to 22 different objects and trash classes. We selected two open-source VLMs for evaluation, namely \blip~\cite{Li2022BLIP} and \clip~\cite{Radford2021CLIP}, based on their availability and applicability to image classification and scene understanding tasks. Both VLMs have shown promising performance in classifying and understanding underwater images in recent studies~\cite{yousafICST2026arxiv,Zheng2024MarineInst,Zheng2023MarineGPT}. 

\subsection{Evaluation Metrics and Statistical Tests} \label{subsec: evaluation_metrics}
\paragraph{\bf Evaluation metrics}
To answer RQ1, we employ the hypervolume (HV) quality indicator. HV measures the volume of the objective space covered by the obtained Pareto front with respect to a reference point. In \ourApproach, it captures the trade-off between minimizing the magnitude of \mr transformations and maximizing the prediction differences. A higher HV value indicates better performance, as it reflects better diversity and convergence toward optimal solutions.

In RQ2, we investigate whether a transformed image causes a prediction failure. To this end, we use a Boolean metric for \mr violations. For \blip, a violation occurs if at least one object count changes in the transformed image compared to the source image. For example, only ``animal'' is predicted in the source image, but ``animal'' and ``vegetation'' are predicted in the transformed image. Given a source image \image and a transformed image \transImage, the violation for \blip is defined as:
\begin{equation}
\BVblip(\image, \transImage) = 
\begin{cases}
1, &\exists\, j \in C \colon
|\countPreTwoArgs{\image}{j} - \countPreTwoArgs{\transImage}{j}| \geq 1 \\
0, & \text{otherwise}
\end{cases}
\label{eq:blip_violation}
\end{equation}

For \clip, a violation is calculated based on the difference between the confidence scores of each class, calculated as:
\begin{equation}
\BVclip(\image, \transImage)= 
\begin{cases}
1, &\exists\, j \in C \colon |\CconfTwoArgs{\image}{j} - \CconfTwoArgs{\transImage}{j}| \geq \thBVclip \\
0, & \text{otherwise}
\end{cases}
\label{eq:clip_violation}
\end{equation}
where \thBVclip is the threshold over which a difference is considered a violation. We experiment with $\thBVclip \in \{1\%, 3\%, 5\%\}$. We compute the violation rate \violRate for each VLM as the ratio of the number of violations to the total number of attempts for a given \mrComposition or order.

Let $\solutionSet = \{(\image_1, \transImage_1), \ldots, (\image_n, \transImage_n)\}$ denote a set of solutions (i.e., source and transformed images). The overall violation rate over \solutionSet is defined as:
\begin{equation}\label{eq:violationRate}
\violRate(\solutionSet) = \frac{1}{|\solutionSet|}\sum_{(\image, \transImage) \in \solutionSet} \BV_{\mathit{vlm}}(\image, \transImage)
\end{equation}
where $\BV_{\mathit{vlm}} (\image,\transImage) \in \{0,1\}$ indicates whether a Boolean violation occurs between source and transformed image for the respective VLM (\blip or \clip) per \Cref{eq:blip_violation,eq:clip_violation}. We will compute the violation rate over two types of solution sets: the set of solutions $\solutionSet_{o}$ obtained with \mrCompositions of a given order ($o\in\{1,$ $\ldots,$ $6\}$); the set of solutions $\solutionSet_{c}$ (with $c\in\{1, \ldots, 64\}$) obtained with a specific \mrComposition among the 64 possible ones.

For RQ3, we measure the violation strength to assess the severity of the detected violations. We employ the mean violation strength metric to quantify the magnitude of violations. For \blip, it represents the average difference in object counts. Namely, given a set \allViolations of pairs of source and transformed images \image and \transImage leading to violations (i.e., $\BV_{\mathit{vlm}}(\image, \transImage) = 1$), the metric is defined as:
\begin{equation}\label{eq:blip_violationStrength}
\begin{array}{l}
\displaystyle
\VSblip(\allViolations) = \frac{1}{|\allViolations|} \sum_{(\image, \transImage) \in \allViolations} \sum_{j \in \overTh(\image, \transImage)} |\countPreTwoArgs{\image}{j} - \countPreTwoArgs{\transImage}{j}|\\
\\
\text{with} \quad \overTh(\image, \transImage) = 
\{j \in C \colon |\countPreTwoArgs{\image}{j} - \countPreTwoArgs{\transImage}{j}| \geq 1\}
\end{array}
\end{equation}
where \overTh represents the set of classes with violations, and 
\VSblip measures the mean violation strength as the average violation magnitude in objects for \blip.
Similarly, for \clip, the mean violation strength represents the average difference in confidence scores at threshold $\thBVclip \in \{1\%, 3\%, 5\%\}$:
\begin{equation}
\label{eq:clip_violationStregth}
\begin{array}{l}
\displaystyle
\VSclip(A) = \frac{1}{|\allViolations|} \sum_{(\image, \transImage) \in \allViolations} \sum_{j \in \overTh(\image, \transImage)} |\CconfTwoArgs{\image}{j} - \CconfTwoArgs{\transImage}{j}|\\
\text{with} \quad \overTh(\image, \transImage) = 
\{j \in C \colon |\CconfTwoArgs{\image}{j} - \CconfTwoArgs{\transImage}{j}| \geq \thBVclip\}
\end{array}
\end{equation}

Higher mean values indicate greater violation strengths and more severe failures, while lower values indicate violations of lower strengths.

\paragraph{\bf Statistical tests}
RQ1 uses HV, whereas RQ3 uses mean violation strength as evaluation metrics. For both RQs, we analyze statistical differences using the Mann–Whitney U test and report effect sizes via Vargha–Delaney (\Atwelve), selected based on the guide in~\cite{guideSBSE}. If the p-value computed by the Mann-Whitney U test is less than 0.05, this indicates a significant difference between \ourApproach and \randomSearch. If there are significant differences, we use \Atwelve to estimate the effect size, categorized in four levels~\cite{kitchenham2017robust}: \neglibleStrength when \(\Atwelve \in (0.44, 0.56)\), \smallStrength when \(\Atwelve \in (0.34, 0.44]\) or \(\Atwelve \in [0.56, 0.64)\), \mediumStrength when \(\Atwelve \in (0.29, 0.34]\) or \(\Atwelve \in [0.64, 0.71)\), \largeStrength when \(\Atwelve \in [0, 0.29]\) or \(\Atwelve \in [0.71, 1]\).

\section{Results and analyses} \label{sec:results}

\subsection{RQ1 -- Comparison with \randomSearch in terms of effectiveness} \label{subsec:RQresults}
\Cref{tab:rq1_summary_hv} reports the results of the statistical tests when comparing \ourApproach, against \randomSearch based on HV values, for \blip and \clip.
\begin{table}[!t]
\centering
\caption{RQ1 -- Comparison of \ourApproach vs \randomSearch for \clip and \blip on 30 images (benchmarks \benchSet) using hypervolume. \Atwelve is reported as \smallStrength, \mediumStrength, or \largeStrength. The best approach is \ourApproach, \randomSearch, or no significant difference (\noDifference)}
\label{tab:rq1_summary_hv}
\begin{subtable}{0.48\textwidth}
\resizebox{\textwidth}{!}{
\begin{tabular}{lccc|ccc}
\toprule
 & \multicolumn{3}{c|}{\clip} & \multicolumn{3}{c}{\blip} \\
\cmidrule(lr){2-4} \cmidrule(lr){5-7}
\benchSet & p-value & \Atwelve & Best & p-value & \Atwelve & Best \\
\midrule
\benchImage{1} & $<$0.001 & \largeStrength & \ourApproach & 0.910 & \none & \noDifference \\
\benchImage{2} & 0.089 & \none & \noDifference & 0.044 & \largeStrength & \randomSearch \\
\benchImage{3} & 0.186 & \none & \noDifference & 0.571 & \none & \noDifference \\
\benchImage{4} & $<$0.001 & \largeStrength & \ourApproach & 0.031 & \largeStrength & \ourApproach \\
\benchImage{5} & $<$0.001 & \largeStrength & \ourApproach & 0.850 & \none & \noDifference \\
\benchImage{6} & 0.076 & \none & \noDifference & 0.002 & \largeStrength & \randomSearch \\
\benchImage{7} & 0.076 & \none & \noDifference & $<$0.001 & \largeStrength & \randomSearch \\
\benchImage{8} & $<$0.001 & \largeStrength & \ourApproach & 0.473 & \none & \noDifference \\
\benchImage{9} & 0.570 & \none & \noDifference & 0.014 & \largeStrength & \ourApproach \\
\benchImage{10} & 0.031 & \largeStrength & \randomSearch & 0.104 & \none & \noDifference \\
\benchImage{11} & $<$0.001 & \largeStrength & \ourApproach & 0.473 & \none & \noDifference \\
\benchImage{12} & 0.089 & \none & \noDifference & 0.021 & \largeStrength & \randomSearch \\
\benchImage{13} & 0.910 & \none & \noDifference & 0.038 & \largeStrength & \ourApproach \\
\benchImage{14} & 0.477 & \none & \noDifference & 0.571 & \none & \noDifference \\
\benchImage{15} & $<$0.001 & \largeStrength & \ourApproach & $<$0.001 & \largeStrength & \ourApproach \\
\bottomrule
\end{tabular}%
}
\end{subtable}
\begin{subtable}{0.48\textwidth}
\resizebox{\textwidth}{!}{
\begin{tabular}{lccc|ccc}
\toprule
& \multicolumn{3}{c|}{\clip} & \multicolumn{3}{c}{\blip} \\
\cmidrule(lr){2-4} \cmidrule(lr){5-7}
\benchSet & p-value & \Atwelve & Best & p-value & \Atwelve & Best \\
\midrule
\benchImage{16} & 0.003 & \largeStrength & \ourApproach & 0.021 & \largeStrength & \ourApproach \\
\benchImage{17} & 0.011 & \largeStrength & \ourApproach & 0.0312 & \largeStrength & \ourApproach \\
\benchImage{18} & 0.677 & \none & \noDifference & 0.009 & \largeStrength & \ourApproach \\
\benchImage{19} & 0.273 & \none & \noDifference & $<$0.001 & \largeStrength & \randomSearch \\
\benchImage{20} & 0.089 & \none & \noDifference & 0.104 & \none & \noDifference \\
\benchImage{21} & 0.004 & \largeStrength & \ourApproach & $<$0.001 & \largeStrength & \ourApproach \\
\benchImage{22} & 0.186 & \none & \noDifference & 0.011 & \largeStrength & \ourApproach \\
\benchImage{23} & $<$0.001 & \largeStrength & \ourApproach & 0.089 & \none & \noDifference \\
\benchImage{24} & 0.007 & \largeStrength & \ourApproach & 0.003 & \largeStrength & \randomSearch \\
\benchImage{25} & 0.850 & \none & \noDifference & 0.002 & \largeStrength & \randomSearch \\
\benchImage{26} & $<$0.001 & \largeStrength & \ourApproach & 0.186 & \none & \noDifference \\
\benchImage{27} & 0.017 & \largeStrength & \ourApproach & 0.006 & \largeStrength & \ourApproach \\
\benchImage{28} & $<$0.001 & \largeStrength & \ourApproach & 0.571 & \none & \noDifference \\
\benchImage{29} & $<$0.001 & \largeStrength & \ourApproach & 0.676 & \none & \noDifference \\
\benchImage{30} & 0.002 & \largeStrength & \ourApproach & 0.021 & \largeStrength & \ourApproach \\
\bottomrule
\end{tabular}%
}
\end{subtable}
\end{table}%
Each row represents a unique image (benchmark image \benchImage{i}). The \textit{Best} column indicates which approach performs significantly better, determined based on the results of the
Whitney U test (p-value column) and \Atwelve value. If no conclusion can be drawn from the statistical test (i.e., $p\text{-value} > 0.05$), the symbol \noDifference is shown to indicate no significant difference between the two compared approaches.

For \clip, the results in \Cref{tab:rq1_summary_hv} show that \randomSearch is significantly better than \ourApproach only once, significantly outperformed by \ourApproach in 16 benchmarks with large effect sizes ($\Atwelve > 0.71$), and no differences in the rest. 
This indicates that \ourApproach consistently achieves equal or superior effectiveness than \randomSearch across the evaluated benchmarks. These results suggest that finding maximum prediction deviations while keeping the transformed images as close as possible to the source images in \clip is challenging, and that a simple algorithm such as \randomSearch is often insufficient, motivating the need for a guided search algorithm such as NSGA-II adopted in \ourApproach.

For \blip, the results in \Cref{tab:rq1_summary_hv} show that: \ourApproach significantly outperforms \randomSearch on 11 benchmarks with large effect sizes, \randomSearch performs significantly better on 7 benchmarks with large effect sizes, and no significant differences are observed for the remaining 12 benchmarks. Overall, these results show that \ourApproach provides a favorable but moderate advantage over \randomSearch for \blip. This suggests that in \blip it is easier to find larger prediction deviations while still preserving high similarity between the transformed and source images. This means that \blip is less reliable than \clip in our context. 

\finding{\textbf{RQ1: } \ourApproach with NSGA-II is effective for testing VLMs, with effectiveness depending on the VLM being tested.}

\subsection{RQ2 -- Binary Violations} \label{subsec:RQ2results}
\paragraph*{\bf RQ2.1 (Analysis by order of \mrCompositions)}
\Cref{tab:rq2_1_MROrders} presents the results of \mrCompositions of different orders (1--6), in terms of percentage of occurrences and binary violation rates \violRate, for \ourApproach and \randomSearch across \clip and \blip.
%
\begin{table}[!t]
\centering
\caption{RQ2.1 -- Analysis by {\it order} of \mrCompositions: occurrence ({\it Occ}) and violation rates (\violRate) of \ourApproach and \randomSearch, for \clip (\Cref{eq:clip_violation}) with three confidence thresholds $\thBVclip \in \{1\%$, $3\%$, $5\%\}$ and for \blip (\Cref{eq:blip_violation}).}
\label{tab:rq2_1_MROrders}
\begin{subtable}{0.60\textwidth}
\centering
\caption{\clip}
\label{tab:rq2_1_clip}
\setlength{\tabcolsep}{2.5pt}
\begin{tabular}{crrrr|rrrr}
\toprule
& \multicolumn{4}{c|}{\textbf{\ourApproach}} & \multicolumn{4}{c}{\textbf{\randomSearch}} \\
\cmidrule(lr){2-5} \cmidrule(lr){6-9}
\ord & $\mathit{Occ}_\%$ & \multicolumn{3}{c|}{\violRatePerc{\thBVclip}} & $\mathit{Occ}_\%$ & \multicolumn{3}{c}{\violRatePerc{\thBVclip}}\\
\cmidrule(lr){3-5} \cmidrule(lr){7-9}
& & 1\% & 3\% & 5\% & & 1\% & 3\% & 5\%\\
\midrule
1 & 26.6\ & 73.3\ & 5.5\ & 0.0\ & 16.7\ & 42.9\ & 1.8\ & 0.0\ \\
2 & 44.2\ & 97.7\ & 39.8\ & 4.6\ & 16.7\ & 63.7\ & 5.2\ & 0.1\ \\
3 & 23.0\ & 99.1\ & 68.1\ & 16.7\ & 16.7\ & 75.2\ & 9.4\ & 0.3\ \\
4 & 5.1\ & 98.7\ & 75.2\ & 22.0\ & 16.7\ & 82.4\ & 13.5\ & 0.6\ \\
5 & 0.8\ & 97.9\ & 69.8\ & 27.9\ & 16.7\ & 87.6\ & 18.2\ & 1.0\ \\
6 & 0.2\ & 99.3\ & 85.8\ & 44.2\ & 16.6\ & 91.7\ & 22.7\ & 1.6\ \\
\bottomrule
\end{tabular}
\end{subtable}
\begin{subtable}{0.39\textwidth}
\centering
\caption{\blip}
\label{tab:rq2_1_blip}
\setlength{\tabcolsep}{2.5pt}
\begin{tabular}{ccc|lc}
\toprule
& \multicolumn{2}{c|}{\textbf{\ourApproach}} & \multicolumn{2}{c}{\textbf{\randomSearch}} \vspace{8pt}\\ 
\cmidrule(lr){2-3} \cmidrule(lr){4-5} 
\ord & $\mathit{Occ}_\%$ & \violRate & $\mathit{Occ}_\%$ & \violRate \vspace{7pt}  \\ 

\midrule 
1 & 74.6\ & 92.8\ & 16.7\ & 37.8\ \\
2 & 17.6\ & 88.1\ & 16.7\ & 47.4\ \\
3 & 5.1\ & 77.8\ & 16.6\ & 51.4\ \\
4 & 2.0\ & 78.3\ & 16.6\ & 54.5\ \\
5 & 0.6\ & 75.2\ & 16.7\ & 56.9\ \\
6 & 0.1\ & 76.7\ & 16.6\ & 58.8\ \\
\bottomrule
\end{tabular}
\end{subtable}
\end{table}%
%
For both VLMs, \randomSearch has roughly equal percentages of \mrCompositions occurrences; each order accounts for around 16.6\%, i.e., it does not target any specific order of \mrComposition. As a result, its violation rate depends mainly on the order; for example, for \clip, \randomSearch achieves violation rates of 42.9\%, 1.8\%, and 0\% at first order for the three thresholds, and increases significantly to 91.7\%, 22.7\%, and 1.6\% at order 6. For \blip, \randomSearch achieves a violation rate of 37.8\% at first order, increasing to 58.8\% at order 6. This indicates that \randomSearch exposes failures mainly via a brute-force approach and high-order \mrCompositions. 
In contrast, \ourApproach, guided by the two optimization objectives, favors lower‑order \mrCompositions. For \clip, \ourApproach favors lower-order \mrCompositions (1-3), each with higher occurrence: 26.6\% for first, 44.2\% for second, and 23\% for third. The higher-order \mrCompositions (i.e., 4--6), instead, have occurrence percentages of 5.1\%, 0.8\%, and 0.2\% respectively. At \violRateOne, \ourApproach achieves violation rates of 73.3\%, 97.7\%, and 99.1\% for lower orders of 1-3; violation rates tend to decrease with higher thresholds, e.g., at 3\% and 5\%, since violations are only considered when differences are large.

For \blip, \ourApproach favors first-and second-order \mrCompositions, resulting in occurrences of 74.6\% and 17.6\%, and yields high violation rates, i.e., 92.8\% and 88.1\%, respectively. Higher-order \mrs (4–6) are rarely selected, collectively <3\% of the total distribution. This implies that \ourApproach, compared to \randomSearch, successfully finds solutions with smaller-order \mrCompositions, resulting in higher occurrences, i.e., producing higher violation rates with smaller transformations on source images for both VLMs.

\paragraph*{\bf RQ2.2 (Top \mrCompositions)}
We investigate which exact \mrCompositions lead to more violations. Tables~\ref{tab:rq2_2_clip} and \ref{tab:rq2_2_blip} provide the top ten \mrCompositions triggering the highest violation rates for \clip and \blip, respectively.
%
\begin{table}[!t]
\centering
\caption{RQ2.2 -- Analysis of top-10 \mrCompositions (\composition) and their violation rates (\violRate) of \ourApproach and \randomSearch, for \clip (\Cref{eq:violationRate}) with three confidence thresholds $\thBVclip \in \{1\%$, $3\%$, $5\%\}$ and for \blip (\Cref{eq:blip_violation}).}
\begin{subtable}{0.75\textwidth}
\centering
\caption{\clip}
\label{tab:rq2_2_clip}
\resizebox{\textwidth}{!}{
\begin{tabular}{lrlr|lrlr|lrlr}
\toprule
\multicolumn{2}{c}{\ourApproach} & \multicolumn{2}{c|}{\randomSearch} & \multicolumn{2}{c}{\ourApproach} & \multicolumn{2}{c|}{\randomSearch} & \multicolumn{2}{c}{\ourApproach} & \multicolumn{2}{c}{\randomSearch} \\ 
\cmidrule(lr){1-2} \cmidrule(lr){3-4} \cmidrule(lr){5-6} \cmidrule(lr){7-8} \cmidrule(lr){9-10} \cmidrule(lr){11-12} 
\composition & \violRatePerc{\thBVclip} & \composition & \violRatePerc{\thBVclip} & \composition &  \violRatePerc{\thBVclip} & \composition &  \violRatePerc{\thBVclip} & \composition &  \violRatePerc{\thBVclip} & \composition &  \violRatePerc{\thBVclip} \\
\cmidrule(lr){2-2} \cmidrule(lr){4-4} \cmidrule(lr){6-6} \cmidrule(lr){8-8} \cmidrule(lr){10-10} \cmidrule(lr){12-12} 
 & 1\% & & 1\% & & 3\% & & 3\% & & 5\% & & 5\%  \\

\midrule

\{2,3\} & 21.6  & \{1,2,3,4,5,6\} & 20.6  & \{2,3\} & 24.8  & \{1,2,3,4,5,6\} & 31.9  & \{1,2,3\} & 20.4  & \{1,2,3,4,5,6\} & 42.8  \\
\{1,3\} & 10.7  & \{1,3,4,5,6\} & 3.5  & \{1,2,3\} & 12.9  & \{1,2,3,5,6\} & 5.4  & \{2,3\} & 15.3  & \{1,2,3,4,5\} & 7.6  \\
\{2\} & 8.5  & \{1,2,3,4,5\} & 3.5  & \{1,3\} & 11.2  & \{1,2,3,4,5\} & 5.3  & \{2,3,5\} & 13.2  & \{1,2,3,5,6\} & 6.7  \\
\{3\} & 7.7  & \{1,2,3,4,6\} & 3.4  & \{2,3,5\} & 10.2  & \{1,2,3,4,6\} & 5.2  & \{1,3\} & 11.8  & \{1,2,3,4,6\} & 6.3  \\
\{1,2,3\} & 6.6  & \{1,2,3,5,6\} & 3.4  & \{1,3,5\} & 5.4  & \{2,3,4,5,6\} & 4.4  & \{1,2,3,5\} & 7.0  & \{2,3,4,5,6\} & 4.0  \\
\{2,3,5\} & 6.1  & \{2,3,4,5,6\} & 3.2  & \{2,3,6\} & 4.5  & \{1,3,4,5,6\} & 4.3  & \{2,3,6\} & 6.7  & \{1,2,3,4\} & 3.1  \\
\{3,5\} & 4.8  & \{3\} & 3.0  & \{1,2,3,5\} & 3.7  & \{1,2,3,6\} & 2.1  & \{1,3,5\} & 3.8  & \{1,3,4,5,6\} & 2.9  \\
\{1\} & 3.8  & \{1,2,4,5,6\} & 2.7  & \{2\} & 2.5  & \{1,2,3,4\} & 2.1  & \{1,2,3,6\} & 3.6  & \{1,2,3,6\} & 2.8  \\
\{1,3,5\} & 3.5  & \{2\} & 2.4  & \{2,6\} & 2.1  & \{1,2,3,5\} & 2.0  & \{1,3,6\} & 3.2  & \{1,2,3,5\} & 2.6  \\
\{2,3,6\} & 2.9  & \{1\} & 2.2  & \{3,5\} & 2.0  & \{2,3,4,5\} & 1.8  & \{2,3,4\} & 2.4  & \{1,2,3\} & 2.4  \\

\bottomrule
\end{tabular}
}
\end{subtable}
\begin{subtable}{0.24\textwidth}
\centering
\caption{\blip}
\label{tab:rq2_2_blip}
\setlength{\tabcolsep}{2pt}
\resizebox{\textwidth}{!}{
\begin{tabular}{lr|lr}
\toprule
\multicolumn{2}{c|}{\ourApproach} & \multicolumn{2}{c}{\randomSearch} \vspace{4pt} \\
\cmidrule(lr){1-2} \cmidrule(lr){3-4} \vspace{4pt}
\composition & \violRate & \composition & \violRate\\
\midrule
\{2\} & 38.1  & \{1,2,3,4,5,6\} & 19.1  \\
\{5\} & 12.4  & \{1,2,3,4,5\} & 3.2  \\
\{1\} & 10.9  & \{1,2,3,5,6\} & 3.2  \\
\{3\} & 7.9  & \{1,2,3,4,6\} & 3.2  \\
\{6\} & 5.9  & \{2,3,4,5,6\} & 3.2  \\
\{2,3\} & 4.2  & \{1,2,4,5,6\} & 3.0  \\
\{3,5\} & 3.3  & \{2\} & 2.9  \\
\{1,3\} & 2.1  & \{1,3,4,5,6\} & 2.9  \\
\{3,6\} & 1.7  & \{6\} & 2.4  \\
\{1,2\} & 1.6  & \{5\} & 2.4  \\
\bottomrule
\end{tabular}
}
\end{subtable}
\end{table}%
For conciseness, we only report the MR number and we omit the prefix ``MR''.

Across both VLMs, \randomSearch produces the most frequent violations with the highest order \mrCompositions (i.e., \{\texttt{1,2,3,4,5,6}\}) obtaining 20.6\%, 31.9\%, 42.8\% of the violations for \clip and 19.1\% for \blip. This \mrComposition produces numerous violations since it significantly alters the image semantics, often resulting in larger changes in target images relative to the source images.

In contrast, \ourApproach identifies lower-order \mrCompositions, which result in violations. For \clip (\Cref{tab:rq2_2_clip}), \ourApproach detects violations that are primarily triggered by \texttt{\mrNum{2}} (Scaling) and \texttt{\mrNum{3}} (Histogram Eq.), e.g., 21.6\% and 24.8\% for \violRateOne and \violRateThree, respectively. The \mrComposition \{\texttt{1,2,3}\} yields the highest number of violations for \violRateFive, i.e., 20.4\%. We observe few single \mrCompositions: \{\texttt{2}\}, \{\texttt{3}\}, \{\texttt{1}\} at \violRateOne, with 8.5\%, 7.7\% and 3.8\% of the violations; and \texttt\{{2}\} at \violRateThree, with 2.5\% violations. Instead, there is no single \mrCompositions for \violRateFive due to the large threshold. These results show that, unlike \randomSearch, \ourApproach discovers more violations while applying fewer \mrs. Moreover, the composition of scale and histogram equalization (\texttt{\mrNum{2},\mrNum{3}}) is particularly effective at revealing VLM failures. \looseness=-1

For \blip (\Cref{tab:rq2_2_blip}), \ourApproach obtains violations largely triggered by single transformations. \mrComposition \{\texttt{2}\} (Scaling) achieves the highest violation rate of 38.1\%, significantly outperforming \randomSearch, which achieves only 2.9\% for the same composition. \mrComposition \{\texttt{5}\} (Shear) and \{\texttt{1}\} (Rotation) rank second and third, achieving the highest violation rates of 12.4\% and 10.9\%, respectively. Pairwise \mrCompositions also produce sufficient violation rates, notably \{\texttt{2,3}\} at 4.2\%. Overall, \blip is more sensitive to single geometric transformations and \ourApproach significantly outperforms \randomSearch in detecting such violations. 

\finding{\textbf{RQ2:} \ourApproach detects violations with fewer transformations than \randomSearch, thereby demonstrating its effectiveness. Moreover, violation effectiveness varies across \mrCompositions and VLMs, showing that optimal compositions are model-dependent.}

%

\subsection{RQ3 -- Strength of Metamorphic Relation Violations} \label{subsec:RQ3results}
\Cref{tab:rq3_strength} reports mean violation strength as defined in \Cref{eq:clip_violationStregth} for \clip (threshold $\thBVclip \in \{1\%, 3\%, 5\%\}$) and in \Cref{eq:blip_violationStrength} for \blip.
\begin{table}[!t]
\centering
\caption{RQ3 -- Comparison by order of \mrCompositions and their mean violation strengths (VS) of \ourApproach and \randomSearch, for \clip (\Cref{eq:clip_violationStregth}) with three confidence thresholds $\thBVclip \in \{1\%$, $3\%$, $5\%\}$ and for \blip (\Cref{eq:blip_violationStrength}). \Atwelve effect size is reported as \smallStrength, \mediumStrength, or \largeStrength, and no significant difference (\noDifference)}
\label{tab:rq3_strength}
\begin{subtable}{0.69\textwidth}
\caption{\clip}
\resizebox{\textwidth}{!}{
\begin{tabular}{cccr|ccr|ccr}
\toprule
\ord & \multicolumn{2}{c}{\violStrengthOne} & \Atwelve & \multicolumn{2}{c}{\violStrengthThree} & \Atwelve & \multicolumn{2}{c}{\violStrengthFive} & \Atwelve\\
\cmidrule{2-3} \cmidrule{5-6} \cmidrule{8-9} 
& \ourApproach & \randomSearch & & \ourApproach & \randomSearch & & \ourApproach & \randomSearch\\
\midrule
1 & 0.05 & 0.03 & \mediumStrength & 0.04 & 0.04 & \noDifference & 0.05 & 0.00 & \noDifference \\
2 & 0.08 & 0.04 & \largeStrength & 0.07 & 0.05 & \mediumStrength & 0.07 & 0.06 & \noDifference \\
3 & 0.11 & 0.04 & \largeStrength & 0.09 & 0.05 & \largeStrength & 0.08 & 0.06 & \mediumStrength \\
4 & 0.12 & 0.04 & \largeStrength & 0.10 & 0.05 & \largeStrength & 0.08 & 0.06 & \largeStrength \\
5 & 0.12 & 0.05 & \largeStrength & 0.11 & 0.05 & \largeStrength & 0.10 & 0.06 & \largeStrength \\
6 & 0.14 & 0.05 & \largeStrength & 0.14 & 0.06 & \largeStrength & 0.09 & 0.06 & \largeStrength \\
\bottomrule
\end{tabular}
}
\end{subtable}
\begin{subtable}{0.28\textwidth}
\caption{\blip}
\label{tab:rq3_blip_severity_summary}
\resizebox{\textwidth}{!}{
\begin{tabular}{cccr}
\toprule
\ord & \multicolumn{2}{c}{\violationStrength} & \Atwelve\\
\cmidrule{2-3} 
& \ourApproach & \randomSearch\\
\midrule
1 & 3.38 & 2.74  & \smallStrength \\
2 & 8.37 & 2.76  & \largeStrength \\
3 & 6.38 & 2.51  & \mediumStrength \\
4 & 3.93 & 2.28  & \mediumStrength \\
5 & 3.95 & 2.25  & \smallStrength \\
6 & 2.87 & 2.21  & \smallStrength \\
\bottomrule
\end{tabular}
}
\end{subtable}
\end{table}
Overall, for \clip, the mean violation strength increases as the order of the \mrCompositions increases; \ourApproach shows lower violation strength at order 1 (0.05, 0.04, and 0.05 for \violStrengthOne, \violStrengthThree, and \violStrengthFive) and the strength increases with increasing order, peaking at 0.14, 0.14, and 0.09 for order 6.\footnote{Only \violStrengthFive shows an exception with a slight decrease from order 5 to order 6.} We also observe a constant mean violation strength at orders 4 and 5 for \violStrengthOne and orders 3 and 4 for \violStrengthFive.

Compared to \ourApproach, \randomSearch shows relatively lower mean violation strengths around (0.03--0.06) across all orders of the \mrCompositions, with a notable value of 0 for order 1.

For \blip, we observe different mean violation strengths across all orders for both approaches. For orders 1-3, \ourApproach shows high violation strength at orders 2 and 3 (8.37 and 6.38, respectively), then dropping to 2.87 at order 6. \randomSearch, instead, shows high strength at orders 1-2 (2.74 and 2.76) with a continuous decrease for the rest of the orders, dropping to 2.21 at order 6.

Overall, \ourApproach, compared to \randomSearch, achieves a greater mean violation strength across all orders. The results of the statistical tests further show that \ourApproach for \clip is significantly better than \randomSearch with medium or large effect sizes in most cases, except for \violStrengthThree in order 1 and for \violStrengthFive in orders 1 and 2, where no significant differences are observed. For \blip, for all orders \ourApproach is significantly better than \randomSearch with small, medium, or large effect sizes.

\finding{\textbf{RQ3:} \ourApproach generally finds violations of significantly higher strengths than \randomSearch, proving its effectiveness in testing VLMs.}

\section{Discussion and Lessons Learned}
\label{sec:discussion}

\textbf{Smaller transformations can reveal failures in VLM-based perception modules.} Our results show that introducing a small number of transformations can cause VLMs to produce incorrect predictions. This suggests that VLM-based perception modules are sensitive to minor changes to inputs, and therefore must be tested for robustness beyond standard benchmark evaluations.

\noindent\textbf{Metamorphic testing provides a practical test oracle.} In the underwater domain, datasets are scarce and ground-truth labeling is limited and expensive. Thus, metamorphic relations offer a scalable way for identifying failures without requiring explicit ground-truth labels.

\noindent\textbf{Search-based testing significantly improves VLM prediction failures discovery over random search.} \ourApproach (NSGA-II) consistently identifies a minimal set of metamorphic transformations that reveal more VLM prediction failures compared to random search, showing that guided exploration of the transformation space is essential for efficient testing.

\noindent\textbf{Different VLMs may need different objective functions.} \blip and \clip, due to their different architectures, produce different types of outputs. We did not use the same objective functions for both models, and as a result, we observed different degrees of violation across the models. This suggests that applying \ourApproach to different VLMs may require different objective functions. However, this does not affect the effectiveness of \ourApproach, as our experiments show it is effective at testing both \clip and \blip.


\section{Threats to Validity} \label{sec:threats}

\noindent
\textbf{Internal validity.} An internal validity threat is related to the configuration of NSGA-II, in \ourApproach, as well as \randomSearch. We chose the default parameter settings across all experiments and compared \ourApproach with \randomSearch to ensure a fair comparison. Another internal threat concerns the order in which active \mrs are applied. In our case, they are applied sequentially in a given order. Different orders may produce different final transformed images and, consequently, may affect the results. This requires additional experiments in the future. 

\noindent
\textbf{External validity.} We consider two generic VLMs and use images selected from the SeaClear dataset. Thus, the results may not generalize to other VLMs or different images. In particular, the considered VLMs differ in their output characteristics, and other models may show different behavior under the same \mr configurations. This highlights the need for further empirical evaluations with additional VLMs, broader datasets, and other perception-related tasks.

\noindent
\textbf{Construct validity.} A construct validity threat is related to the measures used to assess the MR violations. For RQ1, we used hypervolume to evaluate the quality of the obtained Pareto fronts. 
We employed Mann-Whitney U test, p-value, and Vargha-Delaney \Atwelve effect sizes for statistical analysis. For RQ2, we used Boolean violation as a metric and evaluated violation rates. Although these metrics are aligned with the specific output types by VLMs, they remain dependent on the chosen model architectures, i.e., differences in confidence scores and object counts. Another construct validity threat is related to the image selection process. Since we selected 30 benchmark images having no classification errors, it may bias the evaluation towards stable input images. Finally, the set of six \mrs and their parameter ranges were chosen to represent small but meaningful underwater perturbations; different \mrs or different bounds could lead to different results.

\noindent
\textbf{Conclusion validity.} Both \ourApproach and \randomSearch include inherent randomness, which may affect the results. To mitigate this threat, each approach is executed 10 times, following a well-established guide~\cite{guideSBSE}. Moreover, the computational cost of iterative VLM inference limits adding more images and evaluations. To improve confidence in results, we analyzed them with well-established statistical tests as recommended in the same guide.

\section{Related Work}\label{sec:related}


Prior work~\cite{adaptive_mt_testing_OD_models} has used image transformations as \mrs to test deep learning-based systems. Examples include real-world environmental transformations, e.g., blur, rain, and fog~\cite{deepRoad2018}, image-level transformations such as RGB-channel permutation, convolution-order permutation, normalization, scaling~\cite{ml_mt-dwarak2018}, background changes~\cite{deepBackground2021}, object insertion, 3D reconstruction for object detection~\cite{metaod2020}, and scenario-based transformations for multiple object tracking~\cite{mt_dl_multipleObject2022}. Adaptive Metamorphic Testing (MT) has also been used to guide the selection of transformations for image classification~\cite{mt_helg_spieker2020}. 

Few studies have combined search-based and metamorphic testing together to generate test inputs and reveal faults in deep learning models~\cite{mt_deepevolution2019,mt_diverget2022}.
Since exhaustively searching all possible \mrCompositions and their transformation parameters is computationally expensive, we employ multi-objective optimization, i.e., NSGA-II, to better identify \mrCompositions that expose VLM failures. We apply a set of geometric transformations to generate follow-up test images in the underwater domain. While such transformations have been studied in other domains, our novelty lies in applying them to underwater environments and combining multiple transformations simultaneously to generate more realistic transformed images.


MT has been applied to test multimodal systems~\cite{MTsurvey2016}, including image captioning~\cite{mt_imageCaption2022,mt_romeVLM2023}, automated speech recognition~\cite{mt_ASRTest}, and more recently, vision-language action models, visual entailment, and embodied AI~\cite{mt_vlaTest2025,mt_pablo2026,mt_entail2025,mt_metaSpace2026}. In contrast, we focus on testing VLMs in AURs using metamorphic testing, applying multiple transformations to the input image to design effective \mrs.

\section{Conclusion}
\label{sec:conclusion}

This paper proposes a multi-objective, search-based metamorphic testing approach for vision-language models (VLMs) that serve as perception modules for identifying trash in autonomous underwater robots (AURs). We consider two competing objectives: first, to minimize changes to the source underwater images, i.e., to apply as few metamorphic transformations as possible; and second, to maximize changes in the model’s output predictions relative to the source images. We used \ourApproach as a multi-objective search algorithm and \randomSearch search as a baseline. We evaluated two VLMs, namely \clip and \blip. Our results show that \ourApproach is more effective in detecting failures and consistently outperforms \randomSearch by selecting low-order \mrCompositions with higher violation rates. 
Our future work includes studying other specialized VLMs for maritime and multi-objective search algorithms, as well as defining domain-specific metamorphic relations for underwater environments. 

\subsubsection*{Acknowledgments}
This work is supported by the InnoGuard Doctoral Network under the Marie Skłodowska-Curie Actions of the European Commission (Grant Agreement No. 101169233). P. Arcaini is supported by the ASPIRE grant (No. JPMJAP2301) from JST. Aitor Arrieta is a member of the Software and Systems Engineering research group at Mondragon Unibertsitatea (IT1519-22), supported by the Department of Education, Universities and Research of the Basque Country.
\bibliographystyle{splncs04}
\bibliography{references}

\end{document}